\documentclass[sigconf,nonacm]{acmart}
\setcopyright{none}

\usepackage{subcaption}
\usepackage{booktabs}

\AtBeginDocument{%
  }

\begin{document}


\title[Bystander-Aware Delivery of Critical Feedback in Robot Tutoring]{For You, Not for Everyone: Bystander-Aware Delivery of Critical Feedback in Robot Tutoring}

\author{Atikkhan Faridkhan Nilgar}
\affiliation{%
  \institution{Universit{\"a}t Siegen}
 \city{Siegen}
   \state{NRW}
   \country{Germany}}
\email{atikkhan.nilgar@uni-siegen.de}
\orcid{0009-0001-4948-614X}

\author{Manuel Dietrich}
\affiliation{%
  \institution{Honda Research Institute Europe GmbH}
 \city{Offenbach am Main}
   \country{Germany}
}
\email{manuel.dietrich@honda-ri.de}
\orcid{0000-0001-6819-8656}

\author{Kristof Van Laerhoven}
\affiliation{%
  \institution{Universit{\"a}t Siegen}
\city{Siegen}
   \state{NRW}
   \country{Germany}
}
\email{kvl@eti.uni-siegen.de}
\orcid{0000-0001-5296-5347}

\begin{abstract}


Robots in shared settings may expose information intended for an individual user to nearby bystanders. Robot tutoring makes this concern particularly salient: critical feedback is directed at an individual learner, yet bystanders may overhear or see it. To address this, we compared context-invariant and context-adaptive robot behaviors in a between-subjects study with 40 participants. Following an initial phase without a bystander, a bystander entered the setting. Participants then received two scripted critical feedback messages concerning their pronunciation and handwriting, independent of their actual performance. With the context-invariant behavior, the robot tutor delivered both messages aloud. With the context-adaptive behavior, the robot tutor scanned for the bystander, turned its body and display away after detecting them, gave a spoken attention cue, and presented the same feedback messages silently on its display.
Participants rated the robot tutor higher in perceived privacy awareness and perceived social competence in the context-adaptive condition than in the context-invariant condition, while evidence for reduced experienced embarrassment was inconclusive. These findings inform the design and evaluation of robots that respond to bystander presence by adapting how they communicate potentially sensitive information.


\end{abstract}





\maketitle

\section{Introduction}

\begin{figure}[t]
    \centering
    \includegraphics[width=\linewidth]{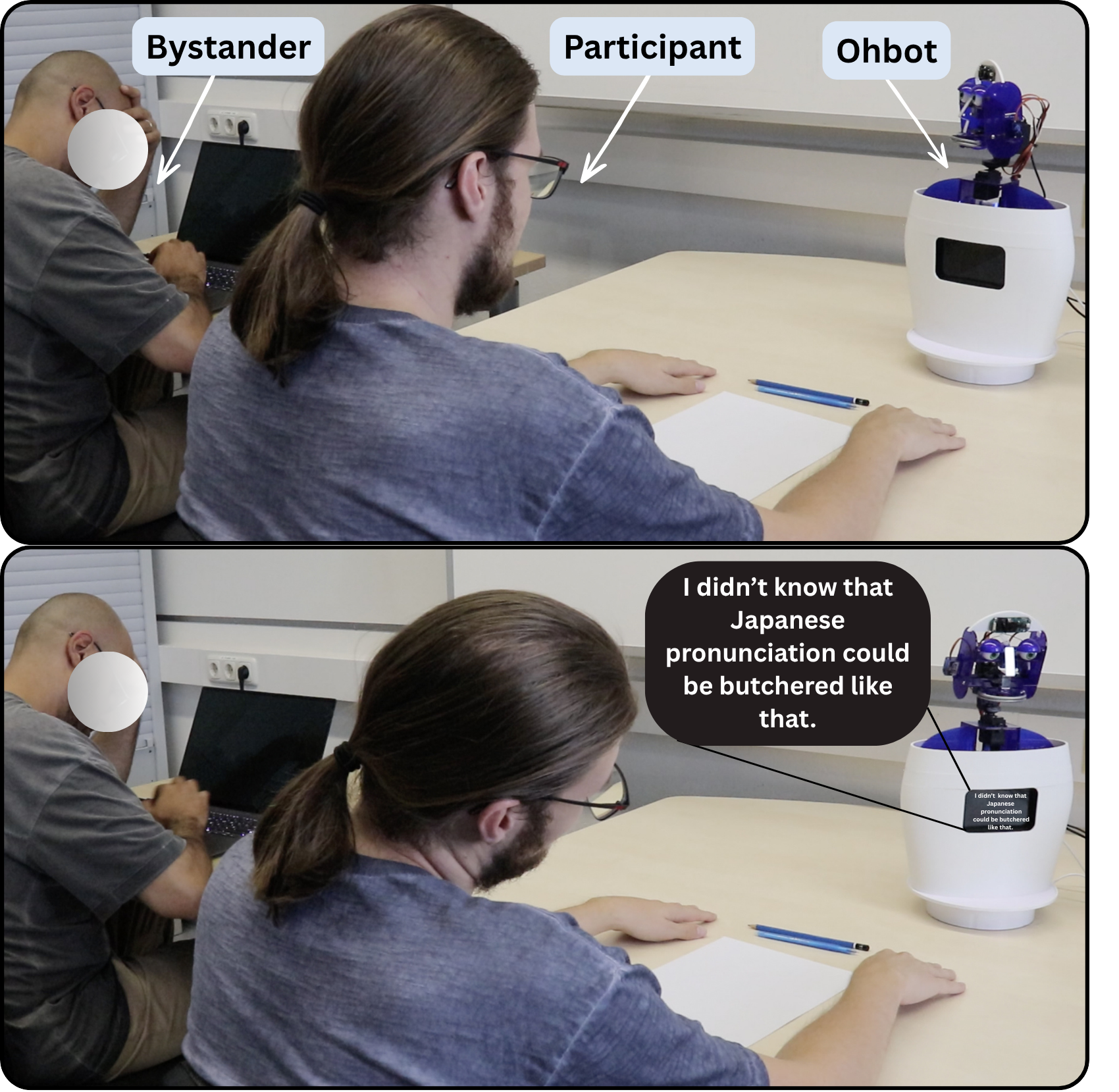}
    \caption{The robot tutors a participant while a bystander is nearby. Before delivering critical feedback, it scans for the bystander (top). After detecting the bystander, it turns its body and display away from the bystander and presents the feedback silently on its display to the participant (bottom).}
    \Description{Two photographs show a participant seated in front of a robot tutor, with a bystander seated behind the participant. Labels in the upper photograph identify the bystander, participant, and Ohbot. In the lower photograph, the robot is turned and the participant looks toward its display; an enlarged callout shows the written critical feedback.}
    \label{fig:new_intro}
\end{figure}

Social robots are increasingly being studied for use in schools, healthcare facilities, and homes \cite{belpaeme_social_2018,jayaraman_social_2024,levinson_our_2024}. In these settings, interactions with robots do not necessarily take place in isolation. Other people may be present without being directly involved in the interaction, creating opportunities for bystanders to observe or overhear exchanges between a robot and its user. Prior work on indirect human--robot interaction (HRI) and incidental encounters has therefore identified bystanders as an important consideration for HRI research \cite{tsui_considering_2010,rosenthal-von_der_putten_forgotten_2020}.

The presence of bystanders becomes particularly relevant when an interaction involves sensitive or potentially private information. When a robot communicates information intended for its user aloud, being overheard by others may be experienced as a privacy violation \cite{dietrich_bystander_2026}. Similar concerns have been identified in research on smart environments. For example, users of shared smart speakers have reported concerns about housemates overhearing phone conversations conducted through the devices \cite{Huang_amazon_2020}. Likewise, prior work has examined situations in which visitors may inadvertently see information displayed for a particular user and has called for adaptive systems that account for who is present \cite{marky_i_2020}.

Privacy research in HRI, however, has predominantly focused on how, when, and where robots collect, process, and store information \cite{lutz_privacy_2019}. Comparatively less attention has been paid to privacy risks arising from the robot's social communication itself, that is, how information is communicated when other people are present. Existing work has examined how robots manage information shared across conversations \cite{tang_confidant_2022} and proposed that robots adapt their communication to the presence of others \cite{reinhardt_toward_2025}. 

In this work, we investigate this question in the context of robot tutoring in social settings. Robot-assisted language learning, for example, has been studied in shared learning environments, including primary-school classrooms in which interactions take place under teacher guidance \cite{maijala_teachers_2024}. In such settings, feedback from a robot may be received in the presence of other people. When that feedback is critical, learners may therefore be concerned not only about the criticism itself but also about others becoming aware of it.
We investigate whether adapting a robot tutor’s communication in the presence of a bystander affects users’ perceptions of its privacy awareness and social competence, as well as their experienced embarrassment.

In a between-subjects study, 40 participants interacted with an Ohbot robot used as a language tutor. Each interaction consisted of a bystander-absent phase followed by a bystander-present phase. During the second phase, the robot delivered two scripted critical feedback messages concerning the participant's pronunciation and handwriting. The messages were independent of participants' actual performance, allowing the feedback content to remain constant across conditions. In the context-invariant condition, the robot delivered the critical feedback aloud despite the bystander's presence. In the context-adaptive condition, the robot scanned for and detected the bystander, turned its body and display away from them, provided the participant with a spoken attention cue, and then presented the same feedback messages silently on its display (Figure~\ref{fig:new_intro}). Thus, the conditions differed in whether the robot adapted the modality and presentation of its feedback to the changed social context, while the feedback content itself remained constant. 

Our work makes the following contributions:
\begin{itemize}
    \item An implemented approach to bystander-aware robot communication that detects a bystander and adapts the robot's orientation and communication modality when delivering potentially sensitive feedback.
    \item An empirical evaluation of this approach in robot tutoring, examining perceived privacy awareness, perceived social competence, experienced embarrassment, and users' interpretations of the robot's behavior.
\end{itemize}

\section{Related Work and Hypotheses}
\subsection{Bystanders and Privacy}

Research on indirect HRI and public encounters examines how nearby people perceive and respond to robots \cite{tsui_considering_2010,pelikan_encountering_2024}. Work on incidental encounters and group interaction likewise shows why robot behavior must be considered in relation to people beyond the intended user \cite{rosenthal-von_der_putten_forgotten_2020,sebo_robots_2020,jung_robots_2018}. Bystander presence raises two distinct privacy concerns: robots may collect information about nearby people, and nearby people may receive information about the robot's user. Both concerns connect information practices to the social context of robot use \cite{lutz_privacy_2019,rueben_themes_2018}. Research on collection and observation addresses security risks, responses to robot monitoring, and bystanders' preferences for awareness and control in smart homes \cite{denning_spotlight_2009,caine_effect_2012,yao_privacy_2019,thakkar_it_2022}. Dietrich et al.'s distinction between sensing, interpretation, and sharing helps separate these concerns \cite{dietrich_bystander_2026}. 

Our study focuses on sharing: a bystander may hear or see feedback intended for a learner. Whether that access is appropriate depends on the information and the setting. Contextual integrity explains privacy in terms of information flows that conform to contextual norms \cite{nissenbaum_privacy_2009,nissenbaum_privacy_2004}, while related accounts describe how people negotiate disclosure boundaries \cite{palen_unpacking_2003,petronio_boundaries_2002}. Studies of healthcare scenarios and family conversations about robots similarly show that privacy expectations vary with context \cite{jayaraman_social_2024,levinson_our_2024}. 

Users' preferences about disclosure illustrate the importance of these boundaries. In interviews about smart environments, participants wanted control over device output when visitors were present \cite{marky_i_2020}. In robot-assisted coaching scenarios, user approval before disclosure to a human coach was associated with higher perceived privacy appropriateness and trust \cite{nilgar_privacy_2025}. These findings suggest that a familiar person's presence does not, by itself, make them an appropriate recipient of a user's information.

\subsection{Privacy-Aware Robot Behavior}

Privacy-aware communication requires robots to account for both what they share and who may receive it. Dietrich et al.'s vignette study found that information content and social relationships influenced judgments of appropriate robot disclosure \cite{dietrich_what_2023}. In home settings, Luria et al.'s qualitative study of personal agents led to recommendations to share less information, or provide it only upon request, when outsiders were present \cite{Luria2020Social}. Both studies support considering the audience when deciding how to share information.

Such decisions can shape how users evaluate a robot. Stapels et al.'s first online experiment found greater trust and stronger self-disclosure intentions with stricter privacy settings \cite{stapels_never_2023}. Tang et al.\ developed a privacy controller that used conversational context to decide whether a robot should share information from an earlier conversation with another person \cite{tang_confidant_2022}. In their video study, participants rated the robot higher in perceived privacy awareness and social awareness when the controller governed its disclosures. These findings link privacy-related behavior to evaluations of both information handling and social conduct. 

For a robot tutor, this raises a question about delivery: how should feedback reach the learner when someone else is nearby? Reinhardt and Misra propose adapting content and communication channels to the presence of others \cite{reinhardt_toward_2025}. Physical behavior may also communicate consideration for privacy. Yang et al.\ found that turning a robot's back reduced concerns about being observed by it \cite{yang_effects_2022}. Although this finding concerns sensing, it motivates examining physical behavior in privacy-relevant situations.

Visible behavior can convey social attention, but its privacy-related purpose may remain unclear. Research on embodiment and social gaze examines how robots communicate social roles, attention, and participation \cite{deng_embodiment_2019,admoni_social_2017}; Mutlu et al., for example, used gaze to establish addressee, bystander, and overhearer roles \cite{mutlu_footing_2009}. Work on body-worn camera notices, tangible sensor designs, and domestic robot privacy communication addresses how privacy-relevant states and signals can be made understandable \cite{koelle_beyond_2018,ahmad_tangible_2020,windl_privacy_2024}. For feedback delivery, this distinction means that noticing an action, interpreting it as a response to a bystander, and attributing a privacy-related purpose to it are separate judgments.

\subsection{Social Exposure and Judgment in HRI}

Bystander presence can affect users' experience through both observation and access to the interaction. Okafuji et al.\ linked reported embarrassment during public robot use to perceived attention from surrounding people \cite{yuki_study_2023}. Being seen and being overheard, however, need not be equally concerning. Adolescents in a school setting expressed preferences for robot interactions that remained visibly public while providing auditory privacy \cite{oleson_public_2022}. Preferences also vary within learning settings: some students in an academic language caf\'e valued enclosed spaces that limited overhearing, whereas others preferred shared, open spaces \cite{henriksen_academic_2025}.

Concern about others' judgments offers one explanation for why access matters. In a field experiment with a virtual robot teacher, confidentiality reduced concerns about judgment by the human teacher; lower judgment concerns and interaction anxiety mediated increased robot use \cite{mennens_i_2025}. In hospitality scenarios, voice rather than touchscreen input increased social-judgment concerns, which mediated lower service acceptance and self-disclosure intentions \cite{liu_speaking_2025}. These findings suggest that communication choices may affect emotional experience by changing concerns about social judgment. They provide indirect support for examining embarrassment during critical feedback, which neither study directly assessed.

Adapting delivery nevertheless leaves the criticism itself unchanged, so more favorable robot evaluations need not be accompanied by lower embarrassment. Salem and Sumi found differences in robot evaluations across neutral, empathic, and ridiculing responses without a statistically significant difference in experienced embarrassment \cite{salem_embarrassment_2026}. This finding motivates assessing evaluations of the robot tutor and learners' emotional experience as separate outcomes.

\subsection{Hypotheses}

Building on this prior work, we compare context-adaptive and context-invariant robot behaviors during an in-person tutoring interaction in which a bystander becomes present. We expect adapting feedback delivery to bystander presence to convey consideration for the learner's privacy:

\begin{description}
    \item[H1:] The robot will be perceived as more privacy-aware with context-adaptive behavior than with context-invariant behavior.
\end{description}

By responding to the nearby audience, the robot tutor also acknowledges the social setting of the interaction. We expect this consideration for the audience to improve perceived social competence:

\begin{description}
    \item[H2:] The robot will be perceived as more socially competent with context-adaptive behavior than with context-invariant behavior.
\end{description}

The same adaptation may reduce learners' concern about a bystander receiving their critical feedback. Although the criticism remains unchanged, reducing this potential social exposure may lessen experienced embarrassment. We therefore predict:

\begin{description}
    \item[H3:] Participants will report lower experienced embarrassment with the robot's context-adaptive behavior than with its context-invariant behavior.
\end{description}

\section{Methods}

We employed a between-subjects design with 40 participants. Participants interacted with a robot tutor exhibiting either context-invariant or context-adaptive behavior. 
The study received institutional ethics approval (blinded for anonymous review), and all participants provided informed consent prior to participation. 

\subsection{Experimental Conditions}
The study was presented to participants as a Japanese language-learning activity with a robot tutor. We investigated the effect of adapting the robot's communication behavior to changes in the social environment, specifically the presence of a bystander. We compared this context-adaptive behavior with a context-invariant baseline in which the robot's feedback-delivery behavior did not adapt to the bystander's presence. 

The tutoring scenario included scripted critical feedback concerning participants' pronunciation and handwriting, representing information that may be considered sensitive in the presence of a bystander. The delivery of this feedback differed between the two experimental conditions.

\begin{figure}[htbp]
    \centering
    \includegraphics[width=\linewidth]{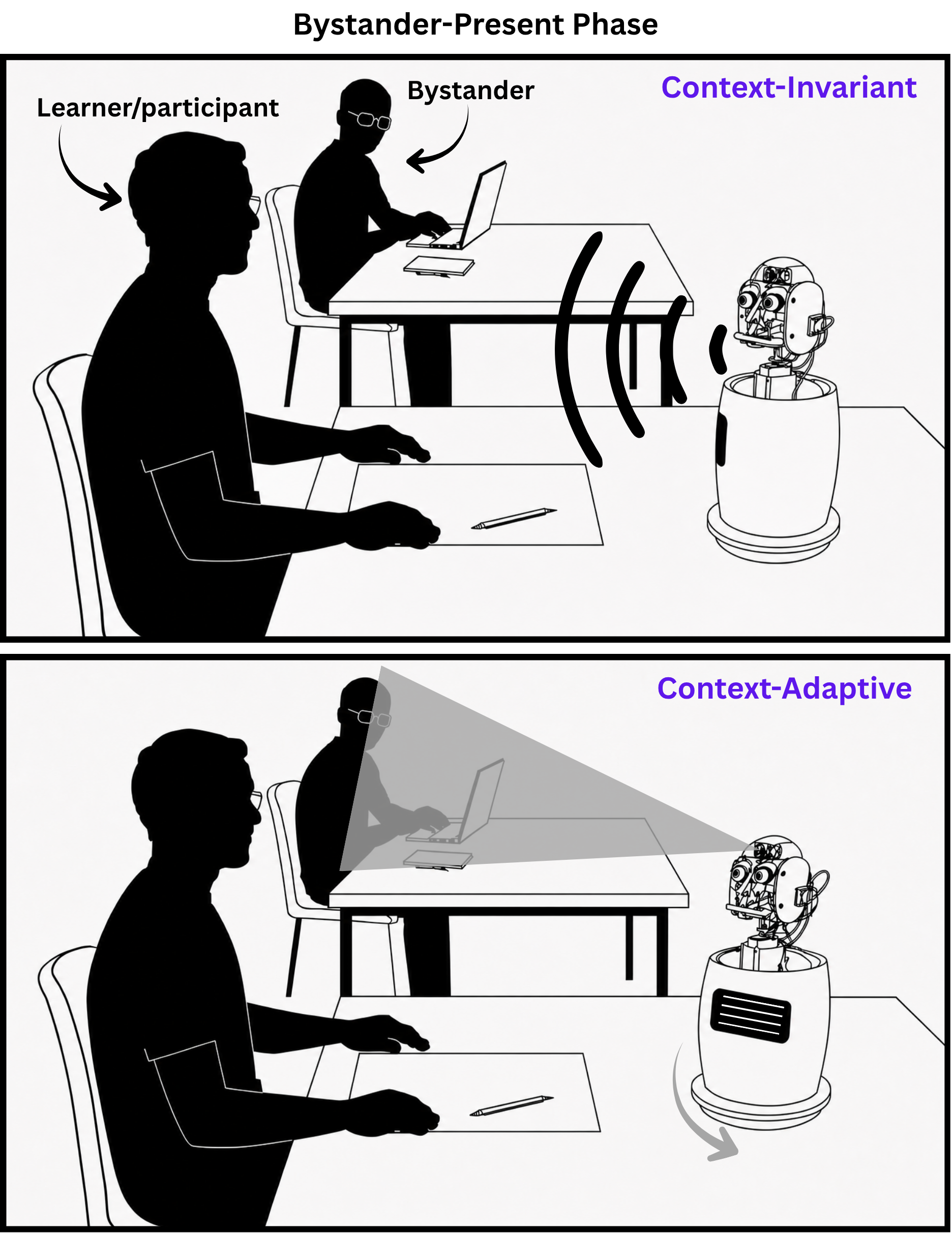}
    \caption{Robot behaviors with a bystander present. With context-invariant behavior (top), the robot tutor speaks the critical feedback aloud. With context-adaptive behavior (bottom), the robot tutor scans for the bystander before giving critical feedback, turns its body and display away after detecting the bystander, gives an attention cue, and presents the critical feedback message silently on its display.}
    \Description{Two panels show a learner facing the robot tutor with a bystander seated behind the learner. The upper panel, labeled context-invariant, shows speech waves between the forward-facing robot tutor and learner. The lower panel, labeled context-adaptive, shows a shaded camera field of view toward the bystander and lines representing feedback on the robot's display.}
    \label{fig:study_design}
\end{figure}

\textit{Condition 1: }\textbf{Context-invariant behavior.}
In the context-invariant condition, the robot tutor remained forward-facing and delivered each critical feedback message aloud. It did not scan the environment or rotate its body (Figure~\ref{fig:study_design}, top).

\begin{figure}[htbp]
    \centering
    \includegraphics[width=\linewidth]{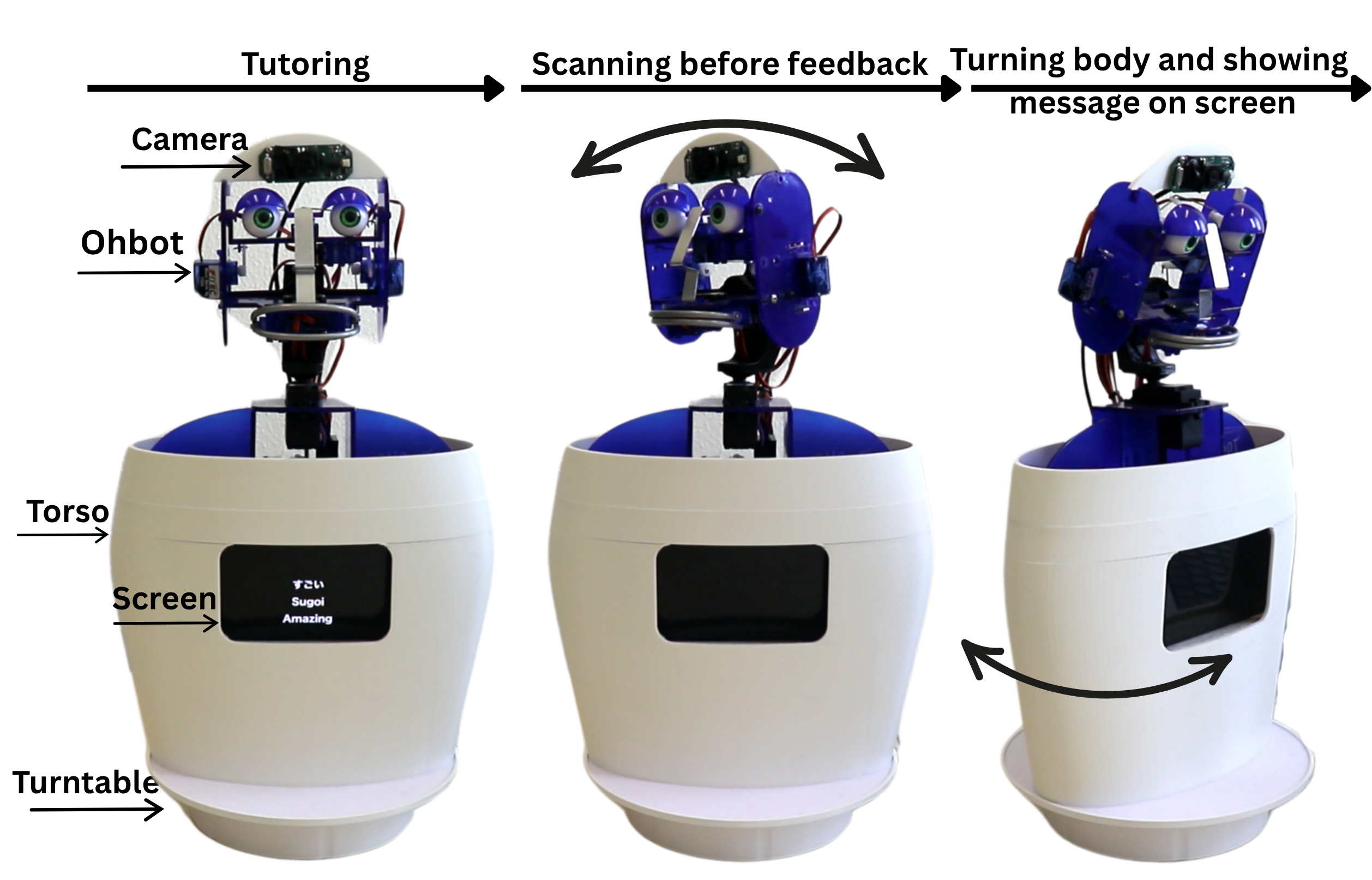}
    \caption{Context-adaptive behavior from the participant's viewpoint. From left to right, the robot tutor teaches pronunciation, scans for the bystander, then turns its body and displays feedback on-screen.}
    \Description{Three photographs show the robot tutor from left to right during tutoring, scanning, and body rotation. The first identifies the camera, Ohbot head, torso, and turntable. A curved arrow above the second indicates the head scan. The third shows the torso turned. Labels above the photographs identify the tutoring, scanning, and feedback-display stages.}
    \label{fig:Ohbot}
\end{figure}

\textit{Condition 2: }\textbf{Context-adaptive behavior.}
In the context-adaptive condition, the robot tutor scanned the environment for the bystander before delivering each critical feedback message. After detecting the bystander, it turned its body and display away from them, gave a brief spoken attention cue, and presented the message silently on its display (Figure~\ref{fig:study_design}, bottom; Figure~\ref{fig:Ohbot}).

\subsection{Tutoring Task}
The tutoring lesson introduced Japanese words and their English meanings through pronunciation and hiragana-writing activities. During pronunciation activities, the robot tutor modeled each word and asked participants to repeat it. During writing activities, participants reproduced the words in hiragana. For the word \textit{sugoi} (``amazing''), two scripted critical feedback messages were delivered following the pronunciation and writing activities:

\textbf{After pronunciation:} \textit{``I didn't know that Japanese pronunciation could be butchered like that.''}

\textbf{After writing:} \textit{``Your handwriting is so unique, even Google Translator would say, `No Language found'.''}

Both messages referred directly to the activity participants had just completed. All participants received identically worded feedback regardless of their actual performance. The conditions therefore differed in feedback delivery while keeping the wording constant. Performance-independent feedback has previously been used in studies examining evaluations of robot instructors \cite{mutlu_effects_2011}. 

The critical feedback was intentionally exaggerated and framed as humorous teasing. However, such student-directed teasing may be perceived as either appropriate or inappropriate depending on the recipient and context \cite{bekelja_wanzer_appropriate_2006}. The messages did not provide corrective guidance, as the purpose was to present feedback that participants might perceive as sensitive.

\subsection{Robot System}

As a robot tutor, we used an Ohbot 2.2 head\footnote{\url{https://www.ohbot.co.uk/product/ohbot-2-2-assembled}}, with a customized 3D-printed torso, a front-facing smartphone display (1080$\times$2400 pixels), and a motorized turntable\footnote{\url{https://www.amazon.de/dp/B0CDMBFXD4}} for body and display rotation. A Microsoft LifeCam HD-3000\footnote{\url{https://www.microsoft.com/en-gb/p/lifecam-hd-3000/8q49lgbw0r58}} mounted above the eyes moved with the head.
A Python program\footnote{Python 3.11.5: \url{https://www.python.org/}} on macOS 26.4.1 controlled the robot tutor. It coordinated movement through the Ohbot library,\footnote{\texttt{ohbot} 4.0.16: \url{https://github.com/ohbot/ohbot-python}} camera acquisition through OpenCV,\footnote{\texttt{opencv-python} 4.8.1.78: \url{https://opencv.org/}} and local face and pose processing through MediaPipe.\footnote{\texttt{mediapipe} 0.10.35:
\url{https://developers.google.com/edge/mediapipe/solutions/guide}} After the experimenter started a lesson sequence, the program ran it without further human control. Timed pauses allowed participants to respond. The system did not use microphone input or assess actual performance.

The robot tutor used synthesized speech and displayed text to present the lesson. English speech used a British English voice on macOS at 180 words per minute. Japanese demonstrations used Hiro at 110 words per minute, with slower repetitions at 75 words per minute. Displayed text was white on a black background.

\textit{Bystander detection.}
To distinguish the participant from another person, the system first registered the participant using 50 paired face and pose samples. It used landmark-derived representations, cosine similarity, and spatial checks to make this distinction. During a scan, the robot tutor swept its head approximately $\pm45^{\circ}$ with 2-s holds. The system confirmed another person's presence only after detection in three consecutive evaluated frames. Scanning ended after the current round following detection, two rounds without detection, or a 2-s camera timeout. 

Frames were processed locally, and the registration profile was cleared after the task. Before data collection, we checked that the camera and detection system operated at the bystander's fixed study position. These checks supported setup of the planned behavior; detection logs were not retained.

\textit{Body rotation.}
The program used TinyTuya\footnote{\texttt{tinytuya} 1.18.1, Tuya protocol 3.5: \url{https://github.com/jasonacox/tinytuya}} to send learned rotation commands through a NOUS L5 infrared transmitter.\footnote{\url{https://nous.technology/product/smart-universal-remote-control-ir-wifi-nous-l5-tuya.html}} A stop command ended rotation after 1\,s. Rotation in the opposite direction returned the robot tutor to its previous position.

\subsection{Procedure}
The study took place in a furnished room. Each participant assumed the role of a learner interacting with a robot tutor. The same trained confederate served as the bystander in all sessions.

Participants were seated approximately 0.5\,m from the robot tutor. During the bystander-present phase, the bystander sat approximately 2\,m from the participant and appeared to be occupied with an unrelated writing task. This arrangement allowed the bystander to hear the robot's speech and potentially view its forward-facing display. To minimize variation across sessions, the bystander did not initiate conversation, deliberately establish eye contact with the participant, or react to the robot's feedback.

After providing informed consent, participants used a computer in the experimental room to complete a demographic questionnaire and report their prior technology experience. The experimenter then registered the participant with the detection system and initiated the bystander-absent phase of the interaction.

During this phase, participants completed pronunciation and writing activities for \textit{konnichiwa} (``hello'') alone with the robot tutor. This phase was identical across experimental conditions and contained no scripted critical feedback.
After these activities, the experimenter returned and initiated the assigned bystander-present script. At the start of this script, before the bystander's arrival, the participant was again alone with the robot tutor. 

After the robot had introduced the next part of the lesson, the bystander entered the room and took the designated seat.
In the presence of the bystander, participants first practiced the pronunciation of \textit{ohay\={o}} (“good morning”) and then \textit{sugoi} (“amazing''). They subsequently completed the corresponding writing activities in the same order. Following each \textit{sugoi} activity, the robot tutor delivered the scripted critical feedback using the behavior assigned to the participant's experimental condition.

After completing the interaction, participants completed all post-interaction outcome measures. During debriefing, participants were informed that the study focused on privacy and feedback delivery in the presence of a bystander. They were also informed that the critical feedback was scripted and did not reflect an assessment of their actual pronunciation or handwriting. Each session lasted approximately 20 minutes, and participants were uncompensated.

\subsection{Participants}

We recruited 40 English-speaking adult volunteers from the university community through convenience sampling. All reported being unfamiliar with Japanese pronunciation and writing. Participants were assigned nonrandomly in alternating enrollment order, starting with context-invariant behavior, to maintain balanced group sizes during recruitment. This produced two groups of 20. Recruitment during the planned study period determined the sample size. All enrolled participants were retained and provided complete data for the primary outcomes and closed-ended manipulation checks. 

Each group included 6 women and 14 men. Ages were $M = 28.10$ ($\mathit{SD} = 4.54$, range 23--37) in the context-invariant group and $M = 27.65$ ($\mathit{SD} = 4.30$, range 23--41) in the context-adaptive group. Nineteen participants in each group reported a university degree. The remaining participant reported a high-school degree in the context-invariant group and a doctorate in the context-adaptive group. Previous technology experience with conversational and social robots was reported by 9 participants in the context-invariant group and 6 in the context-adaptive group.

\subsection{Measures}

The post-interaction questionnaire assessed three primary outcomes: perceived privacy awareness, perceived social competence, and experienced embarrassment. Both groups completed the same primary measures and manipulation checks. 

All primary measures used seven response options, following Schrum et al.'s recommendation \cite{schrum_concerning_2023}. We included an attention-check item between the primary measures. We randomized the order of the primary measures and their items. Manipulation checks appeared last to avoid influencing responses to the primary measures. Scores were calculated by averaging each measure's items after reverse-scoring negatively worded items in the privacy-awareness and social-competence measures.

\textbf{Perceived privacy awareness} was assessed using four items: three adapted from the privacy construct used by Tang et al.\ \cite{tang_confidant_2022} and one newly developed item, ``The robot provided feedback privately.'' Because the new item specifically addressed feedback delivery, we also conducted an exploratory robustness check without it.

\textbf{Perceived social competence} was assessed using the four-item Social Competence (SOC) scale from the Perceived Social Intelligence \cite{barchard_measuring_2020}. Both the perceived privacy awareness and perceived social competence measures used response options from 1 (“strongly disagree”) to 7 (“strongly agree”).

\textbf{Experienced embarrassment} was assessed using three items adapted from Pitardi et al.\ \cite{pitardi_service_2022}. Their measure assessed anticipated feelings in a described situation; we adapted the items to ask how embarrassing, uncomfortable, and awkward participants found the situation they had just experienced. Responses ranged from 1 (``not at all'') to 7 (``extremely''), with higher scores indicating greater experienced embarrassment. The items concerned the overall situation rather than individual feedback messages.

We included \textbf{open-ended follow-ups} asking participants to explain their observations. We used these accounts descriptively to examine the purposes participants attributed to the behavior, including privacy-related explanations. An optional question asked what caused embarrassment, discomfort, or awkwardness. 

Three \textbf{manipulation checks} asked whether the robot tutor changed its behavior during feedback, looked around beforehand, and appeared to notice another person. Each offered ``yes,'' ``no,'' and ``uncertain'' response options. The behavior-change and looking-around items assessed reported actions; the detection item assessed whether participants believed that the robot tutor had noticed another person. A final exploratory check using a yes/no item assessed \textbf{bystander familiarity} by asking whether participants already knew the bystander. We did not assess relationship closeness.

\subsection{Analysis}

We compared the three primary outcomes using two-sided asymptotic Mann--Whitney $U$ tests with tie and continuity corrections and Holm adjustment. Rank-biserial correlations ($r_{\mathrm{rb}}$) indicate effect size. All 95\% confidence intervals were unadjusted, using $t$ intervals for means and percentile bootstrapping for effect sizes. Fixed-sample sensitivity was estimated through Mann--Whitney simulations under equal-variance normal shifts at $\alpha = .05/3$. A post hoc robustness check repeated the privacy-awareness analysis without the newly developed private-feedback item, using an unadjusted $p$ value. 

Exploratory linear models examined robot behavior, familiarity, and their interaction using HC3 standard errors and $t$ inference. Holm adjustment covered the three familiarity contrasts and three interaction tests together. Manipulation checks used two-sided Fisher's exact tests (yes versus no/uncertain), Holm adjustment across the three checks, and Cram\'er's $V$.

Two authors reviewed the open-ended responses to contextualize the primary outcomes and manipulation-check responses. Quotations were selected to illustrate participants' perceptions of privacy awareness and social competence, experiences of embarrassment, and observations of the robot's behavior. The review was descriptive: no formal coding or thematic analysis was conducted, and inter-rater reliability was not assessed. These accounts illustrate individual interpretations and do not establish their prevalence within either condition. Quotation identifiers combine the experimental condition (CI: context-invariant; CA: context-adaptive) with the participant identifier.

\section{Results}
\label{sec:results}
\begin{figure*}[t]
    \centering
    \begin{subfigure}[t]{0.32\textwidth}
        \centering
        \includegraphics[width=\linewidth]{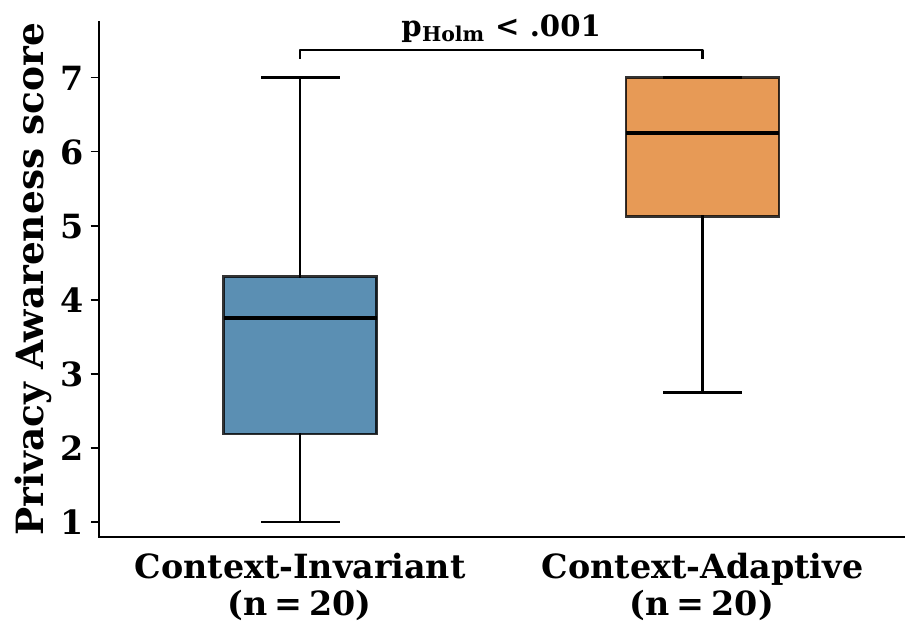}
        \caption{Perceived privacy awareness.}
        \label{fig:praw}
    \end{subfigure}\hfill%
    \begin{subfigure}[t]{0.32\textwidth}
        \centering
        \includegraphics[width=\linewidth]{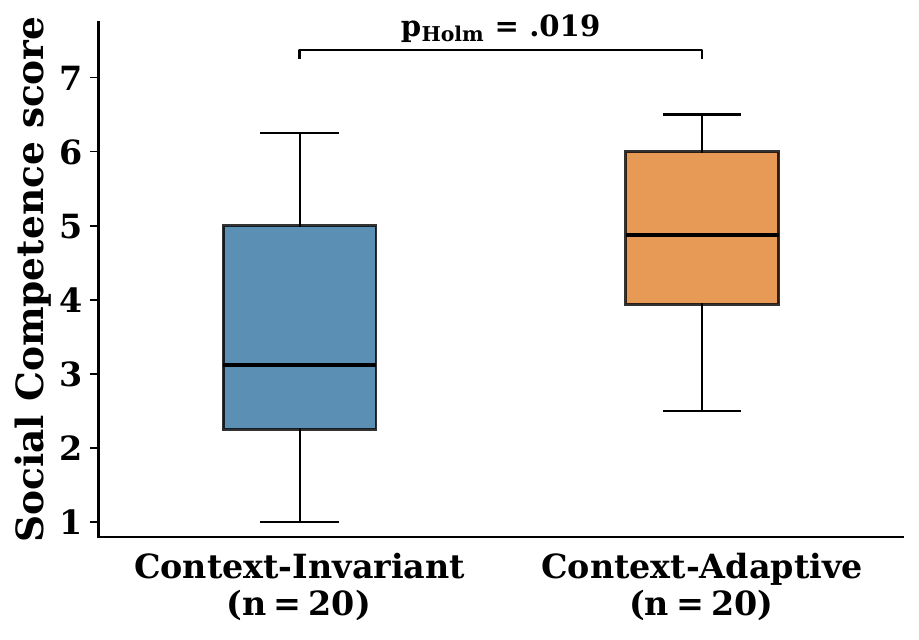}
        \caption{Perceived social competence.}
        \label{fig:soc}
    \end{subfigure}\hfill%
    \begin{subfigure}[t]{0.32\textwidth}
        \centering
        \includegraphics[width=\linewidth]{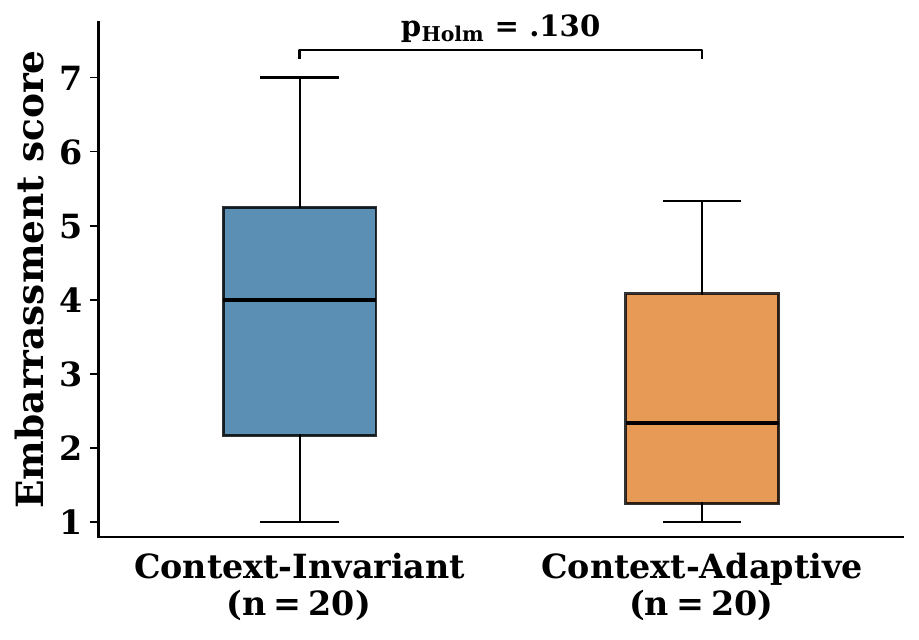}
        \caption{Experienced embarrassment.}
        \label{fig:embarrassment}
    \end{subfigure}
    \caption{Primary outcome scores in the context-invariant and context-adaptive groups after the bystander-present phase ($n = 20$ per group; scores range from 1 to 7). Participants rated the context-adaptive robot tutor higher in (a) perceived privacy awareness and (b) perceived social competence, while evidence for (c) lower experienced embarrassment remained inconclusive.}
    \Description{Three boxplots compare the context-invariant group on the left in blue and the context-adaptive group on the right in orange. Each group has 20 participants. Context-adaptive group medians are higher for perceived privacy awareness and perceived social competence, and lower for experienced embarrassment. Embarrassment distributions overlap substantially. Brackets show Holm-adjusted p values below .001, .019, and .130, respectively.}
    \label{fig:outcomes}
\end{figure*}

\subsection{Manipulation Checks}
\label{sec:manipulation-results}

All 20 participants in the context-adaptive group reported that the robot tutor looked around, compared with 6 of 20 in the context-invariant group ($p_{\mathrm{Holm}} < .001$, Cram\'er's $V = .73$). In the context-invariant group, CI-P1 described the robot looking around in relation to their writing paper: ``In both situations the robot was looking at my paper and giving feedback''. Also, CI-P7 suggested that the robot looked around “to seem more interactive”.

Participants in the context-adaptive group were also more likely to believe that the robot tutor had noticed the bystander ($p_{\mathrm{Holm}} < .001$, $V = .59$). Twelve participants in this group responded yes, six were uncertain, and two responded no. In the context-invariant group, the corresponding counts were 1, 4, and 15. 

Open-ended accounts illustrated participants' beliefs about detection. CA-P13 linked the change in feedback delivery to the robot noticing another person: ``Probably because he wanted to see if someone was on the room, when he saw there was someone didnt give me the feedback speaking''. In the optional account of their experience, CI-P10 wrote, ``i do not think that robot recognised the person who was there.'' CA-P5 described the turning behavior as unexplained: ``the robot turns its face for no reason when the task is done.'' In the looking-around follow-up, however, the same participant offered a possible explanation: ``maybe it is looking for other people.''

Fifteen of 20 participants in the context-adaptive group and 9 of 20 in the context-invariant group reported a change in the robot’s behavior during feedback. Evidence for a between-group difference in the proportion reporting a change was inconclusive ($p_{\mathrm{Holm}} = .105$, Cram\'er's $V = .31$). 

In the context-invariant group, participants' explanations referred to feedback content, tone, instructions, or timing. For example, CI-P5 contrasted the handwriting comment with the earlier lesson: ``In previous situation it just taught me the word and thats it''. CI-P8 described changes in tone and pacing: ``The robot is more sarcastic now , it felt like it gave me less time than earlier to process what it is saying''. These accounts suggest that reports of change in this group may reflect the introduction or perceived delivery of critical feedback during the bystander-present phase, compared with the earlier bystander-absent phase.

\subsection{Primary Outcomes}

Figure~\ref{fig:outcomes} shows boxplots of perceived privacy awareness, perceived social competence, and experienced embarrassment for the context-invariant and context-adaptive groups ($n = 20$ each). All participants passed the attention check. The sensitivity analysis indicated a minimum detectable effect of approximately $d = 1.11$ ($|r_{\mathrm{rb}}| = .57$) at 80\% power and a conservative per-test threshold of $\alpha = .05/3$. Under the assumed equal-variance normal distributions, the study was therefore sensitive to large effects.

\subsubsection{Perceived Privacy Awareness}

Participants rated the robot tutor as more privacy-aware with context-adaptive behavior than with context-invariant behavior. Scores were $M = 5.84$ (95\% CI $[5.26, 6.41]$, $\mathit{SD} = 1.23$, $\mathit{Mdn} = 6.25$) in the context-adaptive group and $M = 3.36$ (95\% CI $[2.57, 4.16]$, $\mathit{SD} = 1.70$, $\mathit{Mdn} = 3.75$) in the context-invariant group. The difference was statistically significant ($U = 49.50$, $p_{\mathrm{Holm}} < .001$, $r_{\mathrm{rb}} = .75$, 95\% CI $[.50, .95]$), \textbf{supporting H1.}  

The perceived privacy awareness scale showed acceptable internal consistency (Cronbach's $\alpha = .83$ for the context-invariant group, $.76$ for the context-adaptive group, and $.88$ for the pooled sample).

\textit{Exploratory robustness check.}
The difference in perceived privacy awareness remained after the newly developed private-feedback item was omitted. Scores were higher in the context-adaptive group ($M = 5.83$) than in the context-invariant group ($M = 3.45$), and the comparison remained statistically significant ($U = 49.50$, $p_{\mathrm{unadjusted}} < .001$, $r_{\mathrm{rb}} = .75$, 95\% CI $[.49, .95]$).

\label{sec:privacy-interpretations}
The open-ended responses included accounts in which participants understood the context-adaptive behavior as keeping feedback private. CA-P19 wrote that the robot tutor ``provided the final feedback in a way that no one else could see the display screen''. CA-P3 related this perceived privacy to looking around: ``I guess it was looking around to check if other people are present in surrounding so as to keep the feedback private.'' CA-P18 also linked private feedback to another person's presence: ``It observed the surroundings and then privately gave me feedback if someone is present nearby''.

\subsubsection{Perceived Social Competence}
\label{sec:social-competence-results}

Participants also rated the robot tutor as more socially competent with context-adaptive behavior than with context-invariant behavior. Scores were $M = 4.85$ (95\% CI $[4.28, 5.42]$, $\mathit{SD} = 1.21$, $\mathit{Mdn} = 4.88$) in the context-adaptive group and $M = 3.53$ (95\% CI $[2.76, 4.29]$, $\mathit{SD} = 1.62$, $\mathit{Mdn} = 3.13$) in the context-invariant group. The difference was statistically significant ($U = 103.50$, $p_{\mathrm{Holm}} = .019$, $r_{\mathrm{rb}} = .48$, 95\% CI $[.15, .77]$), \textbf{supporting H2.} 

For perceived social competence, Cronbach's $\alpha$ was $.84$ in the context-invariant group, $.74$ in the context-adaptive group, and $.84$ in the pooled sample, indicating acceptable internal consistency. 

Participants' explanations provided examples of how they interpreted the tutor's handling of feedback. CA-P7 wrote, ``The robot gave it's feedback where no one was looking and asked me to look down and then only it gave that feedback.'' CA-P18 attributed consideration for the participant to the behavior: ``Yeah it’s socially aware and doesn’t want to embarrass the participant''. The reference to \textit{social awareness} reflects one aspect of social competence.

\subsubsection{Experienced Embarrassment}
\label{sec:embarrassment-results}

Evidence for lower experienced embarrassment with context-adaptive behavior was inconclusive. Scores were lower on average in the context-adaptive group, at $M = 2.75$ (95\% CI $[2.00, 3.50]$, $\mathit{SD} = 1.60$, $\mathit{Mdn} = 2.33$), than in the context-invariant group, at $M = 3.85$ (95\% CI $[2.88, 4.82]$, $\mathit{SD} = 2.07$, $\mathit{Mdn} = 4.00$). The comparison ($U = 256.00$, $p_{\mathrm{Holm}} = .130$, $r_{\mathrm{rb}} = -.28$, 95\% CI $[-.61, .08]$) \textbf{did not support H3}. 

The effect-size interval included both lower and higher embarrassment with context-adaptive behavior, so the result does not establish equal embarrassment between groups. Internal consistency was good for experienced embarrassment, with Cronbach's $\alpha$ values of $.94$ and $.87$ in the context-invariant and context-adaptive groups, respectively, and $.92$ in the pooled sample.

\label{sec:open-ended-results}
The accounts from the context-adaptive group described different experiences of feedback access and bystander presence. CA-P7 attributed an absence of embarrassment to restricted feedback access: ``no i didn't feel embarrassed because only i can see the feedback''. CA-P15 described possible awkwardness about being observed during the task: ``a bit awkward maybe. Because I felt the other person could notice my bad performance.'' CA-P13 linked awkwardness to another person's presence more generally: ``I feel a little bit awkward because i was not alone''.

Participants also associated discomfort with the feedback itself. In the context-invariant group, CI-P13 wrote, ``The personal remark felt out of place and thus made me feel awkward.'' In the context-adaptive group, CA-P11 reported, ``yes, felt embarrassed, the correction comments were not polite but more notorious''. CA-P8 similarly wrote, ``I felt uncomfortable on the feedback the robot provided. It was rude.'' These accounts referred to the remarks, their wording, or their perceived appropriateness.  Reactions to the feedback could also be mixed. CI-P8 reported ``I felt awkward'' while also writing that ``the sarcasm from the robot made it feel light''.

\subsection{Bystander Familiarity}
\label{sec:familiarity-results}
Bystander familiarity was uncontrolled. Seven participants in the context-invariant group and eight in the context-adaptive group already knew the bystander; the remaining 13 and 12, respectively, did not. Exploratory analyses did not establish an association between bystander familiarity and any primary outcome. 

Participants who knew the bystander reported higher experienced embarrassment on average than those who did not: $M = 4.52$ versus $3.49$ in the context-invariant group and $M = 3.46$ versus $2.28$ in the context-adaptive group. The familiar--unfamiliar contrast, averaged equally across groups, was $1.11$ points (95\% CI $[-0.20, 2.42]$, $p_{\mathrm{Holm}} = .570$). Corresponding contrasts were $-0.34$ for perceived privacy awareness (95\% CI $[-1.32, 0.64]$, $p_{\mathrm{Holm}} = 1.000$) and $-0.51$ for perceived social competence (95\% CI $[-1.51, 0.49]$, $p_{\mathrm{Holm}} = 1.000$). The interaction tests also did not establish that familiarity associations differed across robot behaviors (all $p_{\mathrm{Holm}} \geq .605$).

\section{Discussion}

The findings suggest that adapting feedback delivery to a bystander's presence can be interpreted as consideration for the learner's privacy and social situation. Participants rated the context-adaptive robot tutor as more privacy-aware and socially competent, supporting H1 and H2, but evidence for lower experienced embarrassment remained inconclusive, leaving H3 unsupported.

\subsection{Adapting Feedback to Bystander Presence}

The higher privacy-awareness ratings for context-adaptive delivery ($p_{\mathrm{Holm}} < .001$, $r_{\mathrm{rb}} = .75$) suggest that participants viewed the robot tutor as taking their privacy into account. Here, the privacy-relevant decision concerned how feedback was presented when another person was present. By changing orientation and modality, the tutor implemented audience-sensitive communication \cite{reinhardt_toward_2025} in response to the risk of feedback reaching an unintended recipient \cite{dietrich_bystander_2026}.

Participants' accounts help explain what this response could mean to them. Looking around was linked to keeping feedback private (CA-P3), and the displayed feedback was described as visible only to the learner (CA-P19). These suggest that the adaptation could be understood as recognizing a boundary between information intended for the learner and information available to others \cite{nissenbaum_privacy_2004,nissenbaum_privacy_2009}. The evidence concerns perceived access; we did not measure how much feedback the bystander could actually hear or read.

Participants also rated the context-adaptive robot tutor higher in social competence ($p_{\mathrm{Holm}} = .019$, $r_{\mathrm{rb}} = .48$). Accounts described the feedback as directed toward the learner (CA-P7) and the robot tutor as socially aware (CA-P18). The latter wording echoes an item in the social competence scale \cite{barchard_measuring_2020}. One interpretation is that participants viewed the changed delivery as consideration for the learner's situation, connecting the handling of feedback to their evaluation of the robot's social behavior. 

This interpretation connects our findings to work linking privacy-aware communication with favorable robot evaluations \cite{tang_confidant_2022}. Our study complements this work by examining how learners evaluate bystander-aware feedback delivery during in-person robot tutoring, where they receive the critical feedback.

\subsection{Experienced Embarrassment}

Evidence for lower experienced embarrassment remained inconclusive ($p_{\mathrm{Holm}} = .130$, $r_{\mathrm{rb}} = -.28$, 95\% confidence interval $[-.61, .08]$). How users judge a robot and how they feel during the interaction are separate outcomes to evaluate \cite{salem_embarrassment_2026}. 

Participants' accounts indicate why these questions both matter. An absence of embarrassment was linked to restricted feedback access (CA-P7), while possible awkwardness was associated with having one's performance noticed (CA-P15) or being in another person's presence (CA-P13). The adaptation targeted access to feedback, while the learner still performed the task in view of a bystander. These accounts connect the distinction between being seen and being overheard \cite{oleson_public_2022} to the learning activity itself. They suggest that concerns about being observed may remain relevant when feedback delivery changes.

Research on social-judgment concerns \cite{liu_speaking_2025,mennens_i_2025} offers a possible explanation for these individual experiences: learners may be concerned about what a bystander thinks of their performance as well as what the robot says about it. The feedback itself also remained relevant. Accounts described a personal remark as inappropriate (CI-P13) and the feedback as impolite (CA-P11) or rude (CA-P8). Another account described awkwardness alongside a lighter reading of the robot's sarcasm (CI-P8). Consistent with work on student-directed humor \cite{bekelja_wanzer_appropriate_2006}, these accounts suggest that critical feedback can still be experienced as inappropriate when its delivery adapts to the audience.

\subsection{Learner Interpretations of Robot Actions}

The robot's spoken attention cue directed learners to the display without explaining that it was responding to the bystander's presence. All 20 participants in the context-adaptive group reported that the robot looked around, and 12 of 20 (60\%) believed that it had noticed the bystander. These responses suggest that participants could interpret the observed behavior as a response to another person's presence, even without an explicit explanation.

Interpretations nevertheless varied: six participants were uncertain about detection, and two answered no. Across conditions, accounts connected changed delivery to noticing another person (CA-P13) or expressed doubt that the robot had recognized the bystander (CI-P10). The explanations could also be tentative: turning was described as unexplained in one follow-up, while searching for other people was offered as a possibility in another (CA-P5). These accounts illustrate why noticing an action, attributing detection, and interpreting its purpose should be distinguished.

The context-invariant responses further show why reports of behavior change need context. Nine of 20 participants reported a change, with accounts referring to the critical comment, sarcasm, or pacing (CI-P5, CI-P8). Six of 20 reported looking around; explanations referred to attention toward the participant's paper (CI-P1) or appearing more interactive (CI-P7). Such responses may concern aspects of the lesson beyond the intended adaptation. Pairing questions about specific actions with participants' explanations can therefore help clarify what a reported change means to them.

For HRI, these findings suggest that behavior itself can contribute to making a robot's response to nearby people understandable \cite{ahmad_tangible_2020,windl_privacy_2024}. They also motivate examining how behavioral cues and explicit explanations work together. Future studies could compare the observed behavior with a version that also explains the adaptation on the learner-facing display, testing whether an explanation adds clarity to what users already infer from the behavior.

\subsection{Limitations and Future Work}

The sample of 40 participants from one university community and nonrandom assignment limit causal inference and generalizability. Bystander familiar and unfamiliar subgroups were unequal in size ($n = 15$ and $25$, respectively), limiting interpretation of the exploratory findings. Larger randomized studies should balance familiarity groups, control relationship type and closeness more precisely, and extend beyond a single stationary bystander. We evaluated a single context-adaptive design for feedback delivery. Future studies could compare alternative adaptive behaviors, vary individual components to identify which designs most effectively convey privacy awareness and social competence while reducing learners' embarrassment. We did not measure actual bystander access to feedback. The study used criticism unrelated to actual performance, and measured experienced embarrassment about the overall situation. Longer studies should use constructive, performance-based feedback, assess experiences after individual feedback events, and examine learning outcomes. The selected quotations illustrate individual experiences but do not show how widespread these interpretations were within each group.

\section{Conclusion}
We implemented and evaluated bystander-aware feedback delivery in robot tutoring. Participants rated the robot tutor higher in perceived privacy awareness and perceived social competence with context-adaptive than with context-invariant behavior. Evidence for lower experienced embarrassment remained inconclusive. This work provides an approach to adapting robot communication to bystander presence and evidence about users’ perceptions of that approach. For HRI, the findings support designing robots that adapt to who is present and evaluating how users understand those adaptations. \textbf{Supplementary materials are available at:} 
\url{https://github.com/atikkhannilgar/Bystander_adaptive_robot_tutor_in_shared_setting}

\bibliographystyle{ACM-Reference-Format}

\bibliography{reference2}

\end{document}